%% file: kargiantoulakis_qweak_CIPANP2015.tex
\documentclass[12pt]{article}
\usepackage{graphicx}  % For including EPS images
\usepackage{epstopdf}
\usepackage{amssymb} % To add the \square symbol
\usepackage{wrapfig}    % Wrap text around figures
\usepackage[justification=centering,font={small}]{caption} %For centering of captions
\usepackage{cite}
\usepackage{verbatim}

\usepackage{geometry}
\newcommand\pubdate{\today}

\def\uva{University of Virginia, Charlottesville, VA 22908 USA}

\def\Title#1{\begin{center} {\Large #1 } \end{center}}
\def\Author#1{\begin{center}{ \sc #1} \end{center}}
\def\Address#1{\begin{center}{ \it #1} \end{center}}

\newcommand\pubblock{\rightline{%\begin{tabular}{l} \pubnumber\\
         \pubdate}}  %\end{tabular}}}
\newenvironment{Abstract}{\begin{quotation}  }{\end{quotation}}
\newenvironment{Presented}{\begin{quotation} \begin{center} 
             PRESENTED AT\end{center}\bigskip 
      \begin{center}\begin{large}}{\end{large}\end{center} \end{quotation}}
\def\Acknowledgements{\bigskip  \bigskip \begin{center} \begin{large}
             \bf ACKNOWLEDGEMENTS \end{large}\end{center}}
\input econfmacros.tex
\begin{document}
\begin{titlepage}
\pubblock

\vfill
\Title{The $Q_{weak}$ Experiment: First Determination of the Weak Charge of the Proton}
\vfill
\Author{
M.Kargiantoulakis\footnote{e-mail: manos@jlab.org} for the $Q_{weak}$ Collaboration}
\Address{\uva}
\vfill

\begin{Abstract}
The $Q_{weak}$ Collaboration has completed a challenging measurement of the parity-violating asymmetry in elastic electron-proton ($\vec{e}$p) scattering at the Thomas Jefferson National Accelerator Facility (Jefferson Lab). 
The initial result reported here is extracted from the commissioning part of the experiment, constituting about 4\% of the full data set. 
The parity-violating asymmetry at a low momentum transfer $Q^2$=0.025 GeV$^2$ is $A_{ep}$ = -279 $\pm$ 35 (stat) $\pm$ 31 (syst) ppb, which is the smallest and most precise asymmetry ever measured in $\vec{e}$p scattering. This result allowed the first determination of the weak charge of the proton $Q_W^p$ from a global fit of parity-violating elastic scattering (PVES) results from nuclear targets, where earlier data at higher $Q^2$ constrain uncertainties of hadronic structure. 
The value extracted from the global fit is $Q_W^p$ (PVES) = 0.064 $\pm$ 0.012, in agreement with the standard model prediction $Q_W^p$ (SM) = 0.0710 $\pm$ 0.0007. The neutral weak charges of up and down quarks are extracted from a combined fit of the PVES results with a previous atomic parity violation (APV) measurement on $^{133}$Cs. 
The analysis of the full $Q_{weak}$ data is ongoing and expected to yield a value for the asymmetry within 10 ppb of precision.
 Because of the suppression of $Q_W^p$, such a high precision measurement will place significant constraints to models of physics beyond the standard model.
\end{Abstract}
\vfill

\begin{Presented}
Conference on the Intersections of Particle and Nuclear Physics\\
Vail, CO,  May 19--24, 2015
\end{Presented}
\vfill

\end{titlepage}

\def\thefootnote{\fnsymbol{footnote}}
\setcounter{footnote}{0}
%

%\PACSes{\PACSit{12.15}{y}\PACSit{14.20}{Dh}\PACSit{14.65}{Bt} }
% 12.15-y,14.20.Dh,14.65.Bt,25.30.Bf

\section{Introduction}
While the standard model (SM) has been an incredibly successful theoretical framework of particle physics there are many indications of its incompleteness, including neutrino oscillations, dark matter, and the observed matter-antimatter asymmetry. It is considered to be an effective low-energy approximation of a more fundamental underlying structure and searches for physics beyond the SM are well motivated. In the energy frontier, high energy colliders such as the Large Hadron Collider (LHC) are attempting to directly excite matter into new forms. Indirect searches at the intensity frontier offer an important complementary approach through high precision measurements where signatures of physics beyond the SM may appear through quantum loop corrections or tree-level exchange of new particles. \cite{Musolf1994,Erler2005b}

The weak charge of the proton is the vector weak neutral current analog to its electric charge. It is suppressed and precisely predicted in the SM, % (Table~\ref{tab:weak_charges} 
therefore it constitutes an excellent candidate for an indirect probe of new physics \cite{Erler2003}. 
It is connected to the axial electron, vector quark weak couplings $C_{1i} = 2g_{A}^{e}g_{V}^i$ through $Q_W^p = -2(2C_{1u} + C_{1d})$, a combination that is nearly orthogonal to the one accessed by atomic parity violation (APV) experiments. This complementarity allows extraction of the $C_{1i}$ couplings \cite{Young2007} with high precision.

%\begin{table}
%\caption{SM predictions for the weak charges of particles.}
%  \label{tab:weak_charges}
%  \begin{tabular}{rcl}
%    \hline
%      Particle & EM charge & Weak charge \\
%    \hline
%      u       &  2/3   & -2 C$_{1}^u$   \\
%      d       & -1/3   & -2 C$_{1}^d$   \\
%      p (uud) &   1    &   C$_{1}^u$   \\
%      n (udd) &   0    &   C$_{1}^u$   \\
%      
%    \hline
%  \end{tabular}
%\end{table}

The $Q_{weak}$ experiment~\cite{Qweak2007} was performed in experimental Hall C of Jefferson Lab and completed a two year measurement program in May 2012. The experiment measured the parity-violating asymmetry in elastic scattering of electrons from protons in forward angles and low $Q^2$. From the asymmetry the weak charge of the proton $Q_W^p$ 
%=1-4sin^{2}\theta _W$, where $\theta _W$ is the electroweak mixing angle, 
can be extracted and compared to the SM prediction. This comparison can constrain models of new parity-violating physics between electrons and light quarks to the multi-TeV scale. These models include extra neutral gauge bosons, leptoquarks, and parity-violating SUSY interactions~\cite{Erler2003,Musolf2006}. 
The initial results~\cite{Qweak2013} obtained from the analysis of the commissioning run of the experiment are reported here.

The tree level asymmetry can be expressed~\cite{Musolf1994} in terms of Sachs electromagnetic form factors $G_{E}^\gamma$, $G_{M}^\gamma$, weak neutral form factors $G_{E}^Z$, $G_{M}^Z$, and the neutral weak axial form factor $G_{A}^Z$:
\begin{equation}
A_{ep} = \left[
\frac{- G_F Q^2 }{ 4 \pi \alpha \sqrt{2}}
\right]
\left[
\varepsilon G^{\gamma}_{\scriptscriptstyle{E}} G^{Z}_{\scriptscriptstyle{E}} 
+ \tau G^{\gamma}_{\scriptscriptstyle{M}}G^{Z}_{\scriptscriptstyle{M}} 
- (1-4 \sin^2 \theta_W ) \varepsilon^{\prime} G^{\gamma}_{\scriptscriptstyle{M}} G^{Z}_{\scriptscriptstyle{A}}
\over 
\varepsilon (G^{\gamma}_{\scriptscriptstyle{E}})^{2} + \tau (G^{\gamma}_{\scriptscriptstyle{M}})^{2}
\right]
 \label{AepFFs}     
\end{equation}
where $\varepsilon =  \left( 1 + 2(1 + \tau)\tan^2{\theta \over 2} \right) ^{-1} \; \; \rm{and} \; \;
\varepsilon^{\prime} = \sqrt{\tau (1+\tau) (1- \varepsilon^2)}$ are kinematic quantities, $\tau=Q^2$/$4M^2$,  -$Q^2$ is the four-momentum transfer squared, $\alpha$ is the fine structure constant, $M$ is the proton mass, $G_F$ is the Fermi constant and $\theta$ is the laboratory electron scattering angle. 
At forward angles and low momentum transfer it is convenient to recast eq~\ref{AepFFs} as the reduced asymmetry~\cite{Young2007}:
\begin{equation}
{A_{ep}/A_0}  =
   Q_{W}^{p} + Q^2 B (Q^{2},\theta), \;\; A_0=\frac{- G_F Q^2 }{ 4 \pi \alpha \sqrt{2}}
\label{BTermEqn}
\end{equation}

The leading term in eq~\ref{BTermEqn} is the weak charge of the proton which appears as the $Q^2\rightarrow$0 limit of $G_E^Z$. Hadronic uncertainties in terms of electromagnetic, strange and weak form factors enter in the $B(Q^2,\theta)$ term. The low $Q^2$ of the $Q_{weak}$ experiment was chosen specifically to suppress this term without making the asymmetry vanishingly small, so that the high precision goal could be reached within the running period of about two years. 
The hadronic uncertainties can also be constrained from previous PVES experiments at higher $Q^2$ which were aimed at extracting information on the strange and axial content of the nucleon~\cite{Armstrong2012}, 
thus allowing a relatively clean extraction of $Q_W^p$ from the parity-violating asymmetry. At the chosen kinematics the B term contributes about 25\% of the asymmetry.

Beyond tree-level the neutral current couplings are modified by radiative corrections~\cite{Kumar2013}.
The SM radiative corrections for $Q_W^p$ include terms from
ZZ,WW,$\gamma$Z box and other loop diagrams. 
The $\gamma$-Z box diagram $\square_{\gamma Z}^V$(E,Q$^2$) specifically has been the focus of high theoretical interest, %~\cite{Gorchtein2009,Sibirtsev2010,Blunden2011,Rislow2011,Gorchtein2011}. 
when re-analysis of this contribution through forward dispersion relations revealed a significant energy dependence and potentially troublesome theory uncertainty~\cite{Gorchtein2009}. 
Recent analysis however suggests that this contribution is now sufficiently under control~\cite{Hall2013,Rislow2013}. 
It would be desirable to have a unified theory uncertainty on this correction before the full $Q_{weak}$ analysis is completed~\cite{WhitePaper}.

\section{Experimental Overview}

$Q_{weak}$ was performed in experimental Hall C of Jefferson Lab, building on technological advances of previous experiments in the Lab's precision parity violation program \cite{Armstrong2012}. The parameters of the experiment reported here are characteristic of the commissioning phase.
\begin{wrapfigure}{}{0.6\textwidth}
\centering
\includegraphics[width=0.55\textwidth]{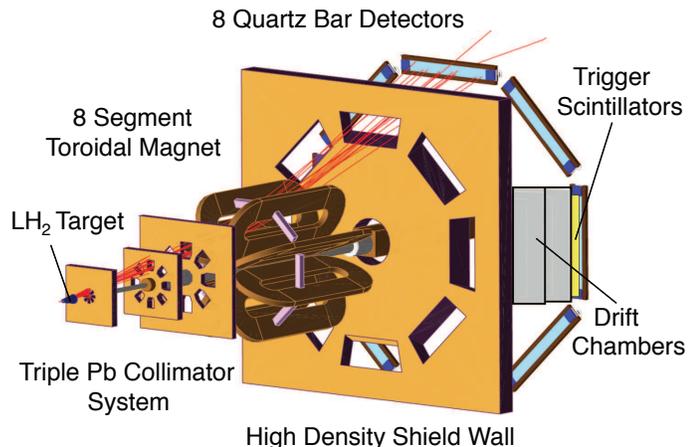}
\captionsetup{width=0.55\textwidth}
\caption{(Color online) The $Q_{weak}$ experimental design. Elastically scattered electrons (red tracks) are being selected by the collimation system and bent by the magnetic field onto the azimuthally symmetric detector system.}
\label{fig:apparatus}
\end{wrapfigure}
The 20 K LH2 cryotarget consisted of a recirculating loop driven by a centrifugal pump, a 3 kW resistive heater, and a 3 kW hybrid heat exchanger making use of both 14 and 4 K helium coolant. It was housed in a 34.4 cm long cell with thin aluminum windows, longer than any previous PVES experiment to maximize interactions and minimize the time needed to achieve the high precision goal. The electron beam at 145 $\mu$A deposited 1.73 kW on the target making this the highest power cryotarget in the world \cite{Smith2012}. 
To reduce beam heating effects on the target the beam was rastered from its intrinsic diameter of $\sim$250 $\mu$m to a 3.5x3.5 mm$^2$ uniform area. 

The experimental apparatus~\cite{Instrumentation2015} is shown in fig~\ref{fig:apparatus}. Upon scattering from the LH2 target electrons at forward angles (7.9$^{\circ}\pm$3$^{\circ}$) are selected by the 3-stage Pb collimator system, which defines the acceptance of the experiment. Heavy shielding is employed to suppress backgrounds in the detector signal. The field of the toroidal resistive dc magnetic spectrometer focuses elastically scattered electrons onto the detector bars while sweeping away inelastics to larger radii. The magnet provided 0.89 T-m at its nominal setting of 8900 A. 

The main detector system consists of an array of 8 radiation-hard synthetic fused quartz (Spectrosil 2000) \u{C}erenkov bars, positioned with azimuthal symmetry around the beam axis at a radius of 3.4 m, 12.2 m downstream of the target. Each detector comprised two 100$\times$18$\times$1.25 cm bars, glued together to form 2 m long bars. The bars were preradiated by 2 cm of lead to suppress soft backgrounds and amplify the electron signal, at the cost of a small increase in the measured asymmetry width. 
Because of the very high luminosity of order 10$^{39}$ s$^{-1}$cm$^{-2}$ and the resulting high detector rate, 640 MHz per detector, the asymmetry data had to be collected in integrating current mode. The \u{C}erenkov light from the detectors was collected by 12.7 cm photomultiplier tubes (PMTs) in either end of the bar assembly. The anode current from the low-gain PMT bases was preamplified by low-noise custom I-to-V converters and then digitized by 18-bit 500 kHz sampling custom-built ADCs.

The experiment was also designed to run in tracking (single pulse) mode at much lower beam currents (0.1-200 nA) for studies of acceptance, $Q^2$ and backgrounds. During these runs a separate PMT base was used and high resolution tracking detectors were inserted around the beamline, including Vertical Drift Chambers (VDCs) and Horizontal Drift Chambers (HDCs) positioned before and after the spectrometer magnet respectively. 

The detector rates are normalized to the beam current as measured by beam current monitors (BCMs). The BCM signal was also used in a feedback loop to suppress charge asymmetry between the two helicity states. 
Because the detector rate also depends on beam parameters such as position, angle, or energy, a helicity-correlated difference in any of these parameters would be the source of a false asymmetry. Orbit differences were continuously measured from beam position monitors (BPMs) upstream of the target, while a BPM in the dispersive region of the accelerator was sensitive to energy differences.

The helicity of the electron beam was controlled by an electro-optic Pockels cell operated at quarter-wave voltage in the Lab's polarized source~\cite{Sinclair2007}. 
The helicity was reversed at 960 Hz, the highest reversal rate ever applied in a PVES experiment, to minimize sensitivity to fluctuations in target density and beam parameters. The detector signal was integrated over each helicity state and the asymmetry was formed from helicity quartets in pseudorandom polarity, (\verb|+--+|) or (\verb|-++-|).  
The laser optics in the polarized source were optimized to minimize helicity-correlated differences in beam parameters that give rise to false asymmetries~\cite{Paschke2009}. A half-wave plate was inserted or removed upstream of the Pockels cell about every 8 hours to reverse the sign of the beam polarization relative to the voltage applied to the cell as a passive cancellation of some classes of false asymmetries. 
Excellent control of helicity-correlated fluctuations in beam parameter properties was achieved through the experiment.

\section{Analysis}

The raw experimental asymmetry is formed as the difference over the sum of the raw charge-normalized detector yields $Y^{\pm}$ in the two helicity states, $A_{raw}$ = ($Y^+$-$Y^-$)/($Y^+$+$Y^-$). The measured asymmetry was extracted after correcting $A_{raw}$ for sources of false asymmetries:
\begin{equation}
A_{msr} =
	  A_{raw}-A_T-A_L-\sum_{i=1}^{5} { \frac {\partial A}{\partial \chi _{i} } \Delta \chi _{i} }
\label{Amsr}
\end{equation}

$A_T$ is the residual transverse polarization on the longitudinally polarized electron beam~\cite{Waidyawansa2013}, 
highly suppressed by the azimuthal symmetry of the detector. It was determined through dedicated measurements with the beam fully transversely polarized, vertically and horizontally. $A_L$ accounts for potential nonlinearities in the PMT response. The last term in eq~\ref{Amsr} is the effect of helicity correlated differences $\Delta \chi _{i}$ in beam orbit or energy, and $\frac {\partial A}{\partial \chi _{i} }$ is the sensitivity of the measured asymmetry to each of these parameters. For the initial $Q_{weak}$ result these sensitivities were extracted from linear regression on natural beam motion. Correction schemes were studied using different sets of BPMs and parameters included in the regression. 
The fully corrected parity-violating asymmetry is obtained after accounting for polarization, backgrounds, and kinematics:
\begin{equation}
A_{ep} =
	  R_{tot} \frac % { \frac {A_{msr}}{P} - \sum_{i=1}^{4} f_i A_i } { 1 - \sum_{i=1}^{4} f_i }
	  	            { A_{msr} / P - \sum_{i=1}^{4} f_i A_i } { 1 - \sum_{i=1}^{4} f_i  } 
\label{Aep}
\end{equation}

The overall factor $R_{tot}$ accounts for the combined effect of radiative corrections, non-uniformities of light and $Q^2$ distribution on the bars, and effective kinematics corrections as in~\cite{HAPPEX2004}. 
The longitudinal beam polarization for the commissioning data set was $P$ = 0.890 $\pm$ 0.018, measured by the M{\o}ller polarimeter~\cite{Hauger2001} in dedicated low current runs.

\begin{wrapfigure}{l}{0.53\textwidth}
\centering
\includegraphics[width=0.50\textwidth]{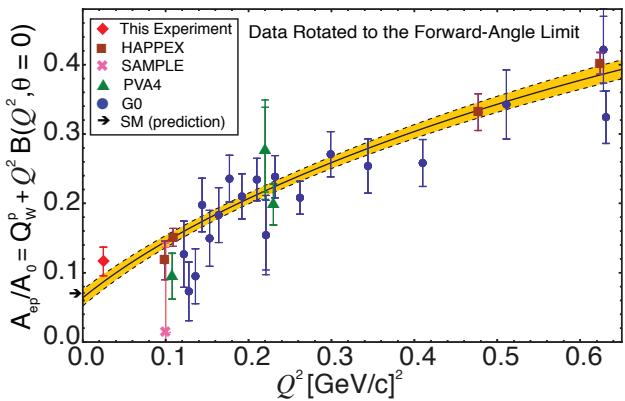}
\captionsetup{width=0.50\textwidth}
\caption{(Color online) Reduced asymmetries (eq~\ref{BTermEqn}) from world PVES on proton up to $Q^2$ = 0.63 (GeV/c)$^2$, including the result of this experiment. The global fit (solid line) includes also data on deuterium and helium. The shaded region is the fit uncertainty, while the intercept is the extracted value for $Q_W^p$, in agreement with the SM prediction (arrow)~\cite{PDG2012}.}
  \label{fig:BtermPlot}
\end{wrapfigure}

The effect of a background source is given as the product of its asymmetry $A_i$ with its dilution $f_i$ (the signal fraction in the main detector). 
The largest background contribution came from the aluminum windows of the target cell. 
The aluminum asymmetry was measured from dedicated runs with dummy targets and the dilution 
was extracted from radiatively corrected measurements with the target cell evacuated. 
The second background correction accounts for scattering from the beamline and the tungsten collimator. The asymmetry and dilution of this source were measured directly by blocking two of the eight openings of the first collimator with 5.1 cm of tungsten. The residual small signal in the blocked main detectors was from this source of background and it was highly correlated to the asymmetries of several background detectors, located outside the acceptance of the main detectors. 
A further correction was applied for the soft neutral backgrounds not accounted for in the blocked octant studies, arising from secondary interactions of electrons scattered in the collimators and magnet. 
The last background correction accounts for inelastically scattered electrons associated with the $N\rightarrow\Delta$ transition. Its asymmetry was directly measured at lower spectrometer magnetic field values and its dilution estimated from simulations.

\section{Results}

For the commissioning run of $Q_{weak}$, comprising about 4\% of the total data, the fully corrected asymmetry  %\cite{Beminiwattha2013} 
from eq~\ref{Aep} is $A_{ep}$ = -279 $\pm$ 35 (stat) $\pm$ 31 (syst) ppb. 
Following the procedure of~\cite{Young2007,Young2006} the weak charge of the proton is extracted from a global fit of PVES asymmetries on hydrogen, deuterium, and $^{4}$He targets, exploiting previous measurements at higher $Q^2$~\cite{SAMPLE2004,SAMPLE2004b,HAPPEX1999,HAPPEX2006,HAPPEX2006b,
HAPPEX2007,HAPPEX2012,G02005,G02010,PVA42004,PVA42005,PVA42009}. 
The Kelly parametrization of electromagnetic form factors~\cite{Kelly2004} 
was adopted and effectively five parameters were free: the weak quark charges $C_{1u}$ and $C_{1d}$, the strange charge radius $\rho _s$ and magnetic moment $\mu _s$, and the isovector axial form factor $G_{A}^{Z(T=1)}$. The value and uncertainty of the isoscalar axial form factor $G_{A}^{Z(T=0)}$ is constrained by the theoretical calculation of~\cite{Zhu2000}.

\begin{wrapfigure}{l}{0.53\textwidth}
\centering
  \includegraphics[width=.50\textwidth]{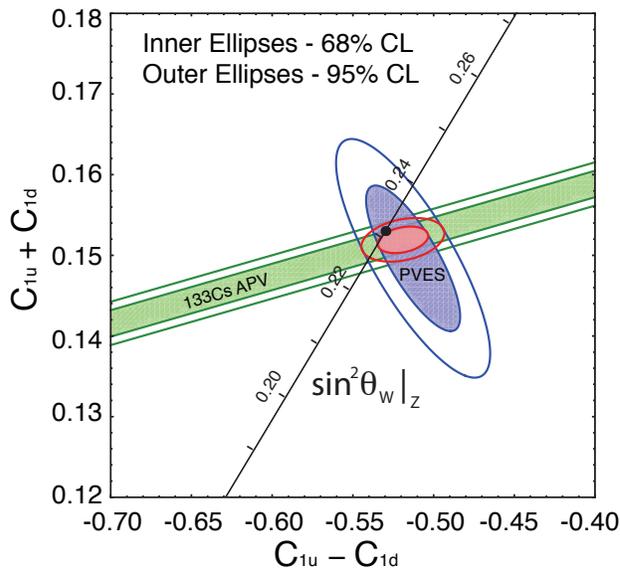}
\captionsetup{width=0.50\textwidth}
  \caption{The constraints on the neutral-weak quark coupling combinations $C_{1u}$ - $C_{1d}$ (isovector) and $C_{1u}$ + $C_{1d}$ (isoscalar) from APV and PVES measurements. The combined global constraints are given by the smaller red ellipse in the center. They are in agreement with the SM prediction (black point), drawn as a function of the weak mixing angle $sin^2\theta_W$ value on the Z-pole~\cite{PDG2012}.
}
  \label{fig:C1_couplings}
\end{wrapfigure}

%% Cut out this part?
The strange quark form factors $G_{E}^s = \rho _s Q^2 G_D$ and $G_{M}^s = \mu _s G_D$ as well as $G_{A}^{Z(T=1)}$ employ a conventional dipole form $G_D$ = (1+$Q^2$/$\lambda^2$)$^{-2}$ with $\lambda$ = 1 (GeV/c) in order to make use of PVES data up to $Q^2 = 0.63$ (GeV/c)$^2$. 
The effects of varying the maximum $Q^2$ or $\theta$ of the data were studied and found to be small above $Q^2  \approx$ 0.25 (GeV/c)$^2$. These four form factors [$G_{E,M}^s$, $G_A^{Z(T = 0,1)}$] have little influence on the results extracted at threshold. The effect of varying the dipole mass in these form factors was studied and found to be small, with a variation of $< \pm$0.001 in $Q_W^p$ for  0.7 (GeV/c)$^2 < \lambda ^2 <$ 2 (GeV/c) $^2$.

All the data used in the fit and shown in fig~\ref{fig:BtermPlot} were individually corrected for the small energy dependence of the $\gamma$-Z box diagram as calculated in~\cite{Hall2013}. 
The even smaller additional correction for the $Q^2$ dependence of the $\square_{\gamma Z}^V$(E,Q$^2$) diagram above $Q^2 = 0.025 (GeV/c)^2$ was included using the prescription provided in~\cite{Gorchtein2011} 
with EM form factors from~\cite{Kelly2004}. 
The small energy and $Q^2$ dependent uncertainties were folded into the systematic error of each point. 

To illustrate the two-dimensional global fit $(\theta , Q^2)$ in a single dimension $(Q^2)$, the angle dependence of the strange and axial form factor contributions was removed by subtracting $[A_{calc} (\theta,Q^2) - A_{calc} (0^{\circ} ,Q^2)]$ from the measured asymmetries $A_{ep}(\theta,Q^2)$, where the calculated asymmetries $A_{calc}$ are determined from eq~\ref{AepFFs} using the results of the fit. The reduced asymmetries from this forward angle rotation of all the ep PVES data used in the global fit are shown in fig~\ref{fig:BtermPlot} along with the result of the fit. The intercept of the fit at $Q^2$=0 is the weak charge of the proton $Q_W^p$ (PVES) = 0.064 $\pm$ 0.012, extracted directly for the first time, in excellent agreement with the SM prediction $Q_W^p$ (SM) = 0.0710 $\pm$ 0.0007.

The fit from PVES data constrains the neutral weak couplings of up and down quarks, with highly complementary sensitivity to the atomic parity violation (APV) $^{133}$Cs result~\cite{APV1997}. A combined fit of the PVES constraints and the APV measurement (fig~\ref{fig:C1_couplings}) yields $C_{1u}$ = -0.1835 $\pm$ 0.0054 and $C_{1d}$ = 0.3355 $\pm$ 0.0050, with a correlation coefficient -0.980. These values for the couplings can be used in turn to obtain a value for $Q_W^p$, $Q_W^p$ (PVES+APV) = -2 (2$C_{1u}$ + $C_{1d}$) = 0.063 $\pm$ 0.012, virtually identical to the result from PVES alone. The $C_1$'s can also be combined to extract the neutron's weak charge $Q_W^n$ (PVES+APV) = -2 ($C_{1u}$ + 2$C_{1d}$) = -0.9890 $\pm$ 0.0007.

\section{Current status}
 
Comparing to the world data set of PVES on nuclear targets, the initial $Q_{weak}$ result constitutes the smallest asymmetry with the smallest absolute uncertainty measured to date. The final result is expected to reach a precision of better than 10 ppb on the parity violating asymmetry.
 A reduction of the systematic uncertainties is necessary to achieve this goal and is the focus of current analysis efforts.

\begin{wrapfigure}{r}{0.6\textwidth}
\centering
  \includegraphics[width=.58\textwidth]{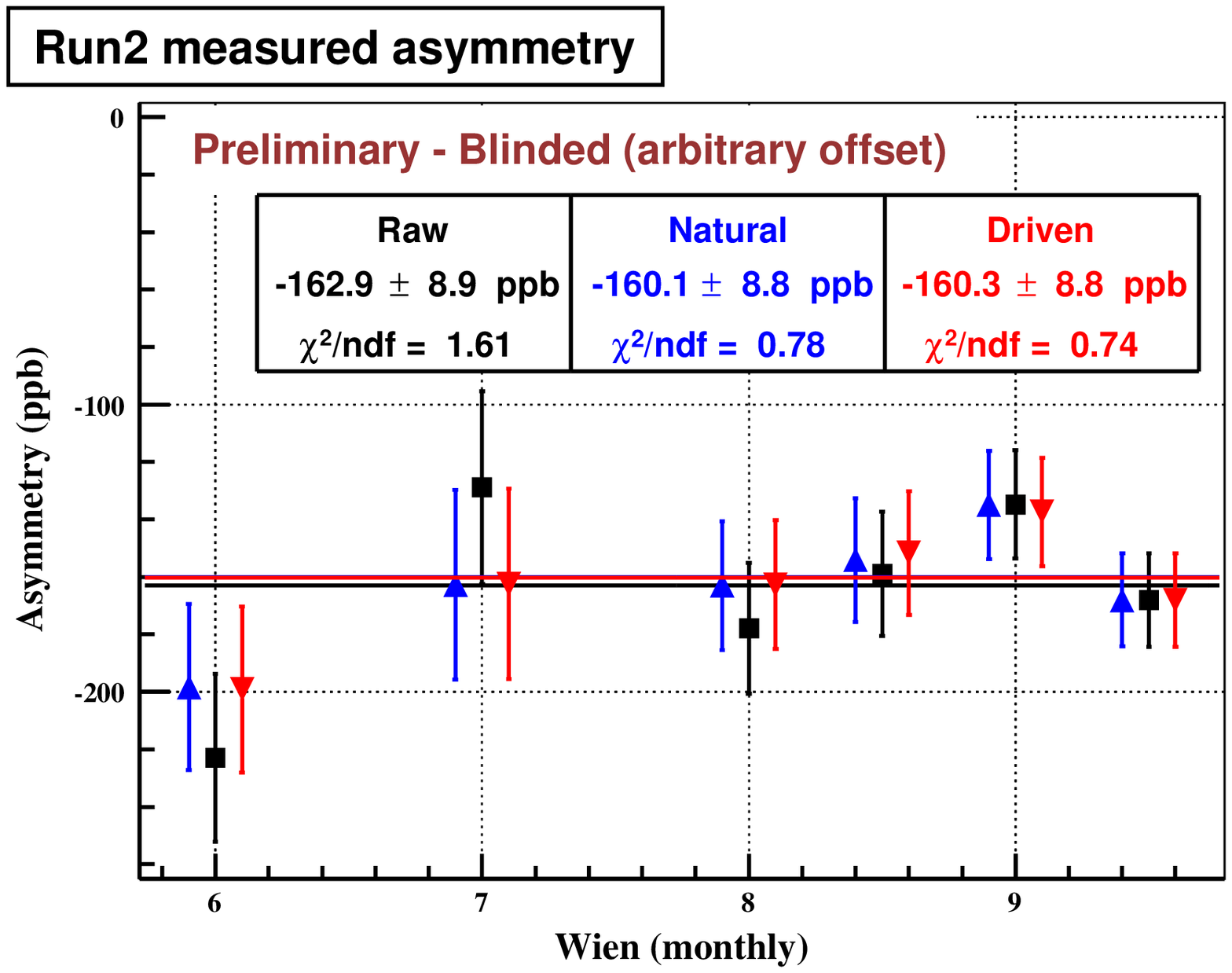}
  \captionsetup{width=0.58\textwidth}
  \captionof{figure}{The regression correction on the raw asymmetry from two independent methods to extract detector sensitivities, from natural and driven motion. 
No other corrections applied to the raw asymmetry, which is also blinded. Approximately half of the full $Q_{weak}$ data set is used in this comparison.}
  \label{fig:Driven_raw}
\end{wrapfigure}

Some important subsystems only became available after the commissioning period of the experiment and were not utilized for the initial result. The Compton polarimeter offered continuous measurements of the polarization concurrent with production and was an important complement to the M{\o}ller. The two redundant polarimeters are preliminarily found to be in very good agreement. 
A new double Wien spin reversal system was installed in the low energy injector to reverse the helicity of the electron beam on the time scale of a month, offering an important cancellation to the effects of higher order helicity correlated differences from the polarized source. 
Another subsystem that became available after commissioning was a set of four air-core dipole magnets in the Hall C beamline and superconducting RF cavities to modulate beam orbit and energy. This procedure allowed an independent measurement of the main detector sensitivities $\frac {\partial A}{\partial \chi _{i} }$ from driven motion. Results from driven and natural motion are preliminarily found to be in good agreement as shown in fig~\ref{fig:Driven_raw}.

The estimated precision of the full $Q_{weak}$ result will significantly constrain models of new physics that would modify the neutral current Langrangian. 
Different prescriptions can be found in the literature for determining the mass reach implied by this result~\cite{Erler2003,Erler2014} in the conventional formalism of contact interactions~\cite{Eichten1983}. 
In the event of a discovery at the LHC, such a high precision measurement will be very important to constrain the characteristics of the new interaction.

%% Old conclusion, referencing Erler2014:
%At a precision of $\sim$5\% of the SM predicted value for $Q_W^p$, the full $Q_{weak}$ result will significantly constrain models of new physics that would modify the neutral current Langrangian. In the conventional formalism of contact interactions~\cite{Eichten1983} the expected reach of the full $Q_{weak}$ result in terms of compositeness scales is $\Lambda >$ 30 TeV, at 95\% confidence level~\cite{Erler2014}. 

\Acknowledgements
This work was supported by DOE Contract No. DE-AC05-06OR23177, under which Jefferson Science Associates, LLC operates. I wish to thank my $Q_{weak}$ collaborators for the opportunity to present on their behalf.

\end{document}

%% file: econfmacros.tex
\def\beq{\begin{equation}}
\def\eeq#1{\label{#1}\end{equation}}
\def\eeqn{\end{equation}}

\def\beqa{\begin{eqnarray}}
\def\eeqa#1{\label{#1}\end{eqnarray}}
\def\eeqan{\end{eqnarray}}

\let\bar=\overbar

\def\Dslash{\not{\hbox{\kern-4pt $D$}}}
\def\dslash{\not{\hbox{\kern-2pt $\del$}}}

\def\msb{{\bar{\ssstyle M \kern -1pt S}}}